\documentclass[12pt]{article}
\pdfoutput=1
\usepackage{amssymb,amsmath,bm}
\usepackage{cite}
\usepackage{graphicx}
\usepackage{longtable,color}

\def\simge{\mathrel{\rlap{\raise 0.511ex \hbox{$>$}}{\lower 0.511ex \hbox{$\sim$}}}}
\def\simle{\mathrel{\rlap{\raise 0.511ex \hbox{$<$}}{\lower 0.511ex \hbox{$\sim$}}}}
\def\slash#1{\setbox0=\hbox{$#1$}\dimen0=\wd0
      \setbox1=\hbox{/} \dimen1=\wd1 \ifdim\dimen0>\dimen1
      \rlap{\hbox to \dimen0{\hfil/\hfil}} #1                        \else
      \rlap{\hbox to \dimen1{\hfil$#1$\hfil}}
      /   \fi}

\newcommand{\newsection}[1]{\section{#1}\setcounter{equation}{0}}

\newcommand{\lsim}{
\mathrel{\hbox{\rlap{\hbox{\lower4pt\hbox{$\sim$}}}\hbox{$<$}}}}

\newcommand{\gsim}{
\mathrel{\hbox{\rlap{\hbox{\lower4pt\hbox{$\sim$}}}\hbox{$>$}}}}

\newcommand{\vtd}{|V_{td}|}

\def\eps{\varepsilon}
\def\epe{\varepsilon'/\varepsilon}
\newcommand{\tev}{\, {\rm TeV}}
\newcommand{\gev}{\, {\rm GeV}}
\newcommand{\mev}{\, {\rm MeV}}

\allowdisplaybreaks[2]

\newcommand{\be}{\begin{equation}}
\newcommand{\ee}{\end{equation}}
\newcommand{\bea}{\begin{eqnarray}}
\newcommand{\eea}{\end{eqnarray}}

\newcommand{\bi}{\begin{itemize}}
\newcommand{\ei}{\end{itemize}}
\newcommand{\ord}{{\cal O}}

\def\kpn{K^+\rightarrow\pi^+\nu\bar\nu}

\def\klpn{K_L\rightarrow\pi^0\nu\bar\nu}

\begin{document}
\begin{titlepage}
\vspace*{-0.5truecm}

\begin{flushright}
TUM-HEP-726/09\\
MPP-2009-68\\
RM3-TH/09-12
\end{flushright}

\vspace{1truecm}

\begin{center}
\boldmath

{\Large\textbf{FCNC Processes in the Littlest Higgs Model \vspace{2mm}\\
with T-Parity: an Update}}

\unboldmath
\end{center}

\vspace{0.4truecm}

\begin{center}
{\bf Monika Blanke$^{a,b}$, Andrzej J.~Buras$^{a,c}$, Bj\"orn Duling$^a$,\\
Stefan Recksiegel$^a$, Cecilia  Tarantino$^d$
}
\vspace{0.4truecm}

{\footnotesize
 {\sl $^a$Physik Department, Technische Universit\"at M\"unchen,
James-Franck-Stra{\ss}e 2, \\D-85748 Garching, Germany}\vspace{0.2truecm}

 {\sl $^b$Max-Planck-Institut f{\"u}r Physik (Werner-Heisenberg-Institut), 
F\"ohringer Ring 6,\\
D-80805 M{\"u}nchen, Germany}\vspace{0.2truecm}

 $^c${\sl TUM Institute for Advanced Study, Technische Universit\"at M\"unchen,  Arcisstr.~21,\\ D-80333 M\"unchen, Germany}\vspace{0.2cm}

 $^d${\sl Dipartimento di Fisica, Universit\`a di Roma Tre and INFN, Sez. di Roma Tre,\vspace{-1mm}\\
 Via della Vasca Navale 84, I-00146 Roma, Italy}
}

\end{center}

\vspace{0.6cm}
\begin{abstract}
\noindent
We update our 2006-2007 results for FCNC processes in the Littlest Higgs
 model with T-parity (LHT). The removal of the logarithmic UV cutoff dependence in
our previous results through a new contribution to the $Z^0$-penguin diagrams 
identified by Goto et al.\ and del Aguila et al., while making the deviations from
the SM expectations in the quark sector less spectacular, still allows 
for sizable {new physics effects} in $K\to\pi\nu\bar\nu$ and $K_L\to \pi^0 \ell^+\ell^-$ decays and in 
the CP-asymmetry $S_{\psi\phi}$ with the latter unaffected by the new contribution. We extend our analysis {by a study of the fine-tuning required to fit the data on $\eps_K$} and {by the inclusion of} the decay $K_L\to\mu^+\mu^-$. A number of correlations can distinguish this model from the custodially protected {Randall-Sundrum model} analysed recently.
We also reconsider lepton flavour violating decays, including now a discussion of fine-tuning. 
While the $\ell_i\to \ell_j\gamma$ decays are 
unaffected by {the removal of the logarithmic cutoff dependence}, the branching ratios for decays with three
leptons in the final state, like $\mu\to 3 e$ are lowered by almost an
order of magnitude. In spite of this, the pattern of lepton flavour violation in
the LHT model can still be distinguished from the one in supersymmetric
models. 
\end{abstract}

\end{titlepage}
\setcounter{page}{1}
\pagenumbering{arabic}

\newsection{Introduction}
The most extensive studies of FCNC processes beyond the framework of minimal
flavour violation (MFV) \cite{Buras:2000dm,Buras:2003jf,D'Ambrosio:2002ex,Chivukula:1987py,Hall:1990ac} {are at present} performed in models that address the
question of stability of the Higgs mass under radiative corrections.
The most popular approaches in this context are the general
MSSM (GMSSM) \cite{Martin:1997ns,Peskin:2008nw,:2008gva}, Randall-Sundrum (RS) models \cite{Randall:1999ee} with bulk fields \cite{Davoudiasl:1999tf,Pomarol:1999ad,Chang:1999nh,Grossman:1999ra,Gherghetta:2000qt,Contino:2006qr,Cacciapaglia:2006gp,Carena:2006bn,Casagrande:2008hr,Albrecht:2009xr} and the Littlest Higgs model \cite{ArkaniHamed:2001ca,ArkaniHamed:2001nc,Schmaltz:2005ky,Perelstein:2005ka,ArkaniHamed:2002qy} with
T-parity (LHT) \cite{Cheng:2003ju,Cheng:2004yc,Low:2004xc}. In these models large, or even spectacular deviations from the 
Standard Model (SM) predictions are still possible while satisfying all
existing constraints from quark flavour and lepton flavour violating
processes.

As flavour physics is entering the era of precision tests and the LHC will
soon provide new data on very high energy collisions it is realistic to
expect that within the next decade we will find out whether any of these
New Physics (NP) scenarios is chosen by nature and whether large deviations
from the SM and MFV  expectations as predicted in these models are observed in the
data.

Among the three non-MFV scenarios in question the LHT model is rather
special for the following reasons:
\begin{itemize}
\item
It contains the smallest number of new parameters in the flavour sector: 10
in the quark sector \cite{Blanke:2006xr} to be compared to 63 in the GMSSM and 27 in the simplest 
RS models with bulk fermions \cite{Agashe:2004cp}.
\item
Only SM operators are relevant in this model \cite{Blanke:2006sb,Blanke:2006eb} in contrast to the GMSSM and RS
models in which new effective operators can 
have significant impact and bring in
new hadronic uncertainties \cite{Bona:2007vi,Csaki:2008zd,Blanke:2008zb,Blanke:2008yr}.
\item
The LHT scale can be as low as $f=500\gev$ \cite{Hubisz:2005tx}, in contrast to RS models, where
electroweak (EW) precision tests even in models with custodial protection \cite{Agashe:2003zs,Csaki:2003zu,Agashe:2004rs,Agashe:2006at}
require $M_\text{KK}\ge (2-3)\tev$ \cite{Bouchart:2008vp,Djouadi:2006rk}.
\item
The constraints from $B\to X_s\gamma$ and $d_n$ are easily satisfied \cite{Blanke:2006sb}, while these observables
put severe constraints on the GMSSM and RS models \cite{Agashe:2004cp,Agashe:2008uz}.
\item
New sources of flavour and CP violation in the quark sector are fully encoded in two 
$3\times 3$ unitary matrices $V_{Hd}$ and $V_{Hu}$ that are related to 
each other
through $V_{Hu}=V_{Hd} V_{\rm CKM}^\dagger$ \cite{Hubisz:2005bd,Blanke:2006xr},
implying thereby correlations between FCNC processes in the down-quark
and up-quark sectors \cite{Blum:2009sk,Bigi:2009df}.
\end{itemize}

In a series of papers we have performed an extensive study of 
particle-antiparticle mixing \cite{Blanke:2006sb,Blanke:2008ac}, 
rare $K$ and $B$ decays \cite{Blanke:2006eb},
lepton flavour violating (LFV) processes \cite{Blanke:2007db}, $\epe$ 
\cite{Blanke:2007wr}  and 
$D^0-\bar D^0$ mixing  
\cite{Blanke:2007ee,Bigi:2009df}. Selected reviews of these investigations can be found in 
\cite{Tarantino:2007kg,Blanke:2007ww,Duling:2007sf}.

The goal of the present paper is an update of our results for 
rare $K$ and $B$ decays and LFV processes. $D^0-\bar D^0$
mixing and CP violation in $D$ decays have been considered in a separate
publication \cite{Bigi:2009df}. The main reason for this update is the fact that in our 
previous papers we have overlooked an $\mathcal{O}(v^2/f^2)$ contribution
to {the} $Z^0$-penguin diagrams.
This contribution has been identified by Goto et al. 
\cite{Goto:2008fj}
in the context of their study of {the} $K\to\pi\nu\bar\nu$ decays in the LHT model,
and independently by del Aguila et al. \cite{delAguila:2008zu} in the context  of the corresponding analysis
of the LFV decays $\mu\to e\gamma$ and $\mu\to3e$. At the same time, these authors have confirmed our calculations except for the omission mentioned above.

The main virtue of the new contribution is the cancellation of the logarithmic
UV cutoff dependence present in our results. Such cutoff dependence is 
characteristic for models like the LHT in which the UV completion has not been
specified, and in fact such a sensitivity is present in the LH model without
T parity \cite{Buras:2006wk}.
However, it turns out
that in the LHT model the logarithmic UV cutoff dependence is absent at
$\mathcal{O}(v^2/f^2)$ making the predictions in this model much less sensitive
to the physics at the UV cutoff scale of this model.

It should be emphasised at this stage that all the decays considered by
us that do not receive $Z^0$-penguin contributions are unaffected by
these modifications. Thus our formulae for particle-antiparticle mixing
observables and branching ratios for $B\to X_s\gamma$ and $\ell_i\to \ell_j\gamma$
decays remain intact.

Our paper is organised as follows. In Section \ref{sec:cutoff} we {summarise  briefly what is new in our paper.
The corrected Feynman rules of} \cite{Blanke:2006eb} implied by the findings of \cite{Goto:2008fj,delAguila:2008zu} are collected in Appendix \ref{sec:appA}.
In Section \ref{sec:KLmumu} we present the formula for LHT contributions to the decay $K_L\to\mu^+\mu^-$ that we did not consider in our previous papers. In Section \ref{sec:num}
we collect the input parameters and describe our strategy for the
numerical analysis. In Section \ref{sec:KB} the new results for the quark sector
are presented. The corresponding results for LFV
processes are collected in Section \ref{sec:LFV}, extended by a discussion of the necessary fine-tuning in $\mu\to e\gamma$ and $\mu\to3e$. We close our paper with a brief
summary of our results and with an outlook.

\newsection{What's New in Our Analysis}\label{sec:cutoff}
On the theoretical side, as already advertised in the Introduction, in our
previous papers that involved $Z^0$-penguin contributions, we have omitted 
an $\ord(v^2/f^2)$ correction to some of the vertices with right-handed
 mirror fermions. {The Feynman rules for the couplings of mirror fermions to SM gauge bosons in \cite{Blanke:2006eb} have been derived assuming strictly vectorlike couplings. While this assumption is viable at leading order -- otherwise EW symmetry would be broken explicitly -- a possible $\ord(v^2/f^2)$ correction does not need to {fulfil} this requirement and has thus been overlooked.} This correction has first been pointed out by Goto et al.\
\cite{Goto:2008fj} and subsequently by del Aguila et al.\ \cite{delAguila:2008zu}, and we confirm their result. As these authors discuss 
the origin of this additional term in detail, we will not repeat this 
discussion here.
 The corrected Feynman rules of 
\cite{Blanke:2006eb} implied by the findings of 
\cite{Goto:2008fj,delAguila:2008zu} are collected in Appendix \ref{sec:appA}.

In the process of updating our analysis we have also modified the $U(1)_i$ charges of the lepton multiplets in agreement with the charge assignments adopted in  \cite{Goto:2008fj}. {While the lepton charge assignments chosen in \cite{Hubisz:2004ft} and adopted by us in \cite{Blanke:2006eb} are conceptually viable, for phenomenological purposes the assignments of \cite{Goto:2008fj} are more convenient since they allow to implement the lepton sector in complete analogy to the quark sector. We stress that this change of charge assignments is completely unrelated to the omitted $\ord(v^2/f^2)$ correction in some of the mirror fermion couplings. {We confirm the finding of \cite{Goto:2008fj} that the impact of this change is numerically irrelevant.}}

Concerning phenomenology, in addition to the update of the most interesting 
observables, we will analyse the degree of fine-tuning necessary to satisfy
the {constraints from} $\varepsilon_K$ and the $\mu\to e\gamma$ decay.
We also extend our analysis to include the decay $K_L\to\mu^+\mu^-$ which exhibits
an interesting correlation with $K^+\to\pi^+\nu\bar\nu$ that differs from
the one found in the RS model with custodial protection \cite{Blanke:2008yr}.

{Throughout this paper we use the notations and conventions of our previous LHT
analyses. For a detailed model description we refer the reader to
\cite{Blanke:2006eb}. The formulae used by us are the ones contained in
our previous papers on the LHT model modified appropriately as outlined in the Appendix.}

\boldmath
\newsection{$K_L\to\mu^+\mu^-$ in the LHT Model}\label{sec:KLmumu}
\unboldmath

In the present analysis {we add} the $K_L\to\mu^+\mu^-$ decay to the discussion of {rare decays} in the LHT model. This mode offers another interesting possibility to probe the NP flavour structure. Analytic expressions for the short distance (SD) contribution can easily be obtained from 
Section 4.4 of \cite{Blanke:2008yr}.

In the LHT model, following \cite{Buras:2004ub,Blanke:2008yr}, we thus 
have ($\lambda=0.226$)
\be
Br(K_L\to\mu^+\mu^-)_{\rm SD} =
 2.08\cdot 10^{-9} \left[\bar P_c\left(Y_K\right)+
A^2 R_t\left|{Y_K}\right|\cos\beta_{Y}^K\right]^2\,.
\ee
The expression for the loop function $Y_K\equiv|Y_K|e^{i\theta_Y^K}$ can be obtained from (3.4) of \cite{Blanke:2006eb} by introducing the correction outlined in Appendix \ref{sec:appA}. In addition we have defined
\begin{gather}
\beta_{Y}^K \equiv \beta-\beta_s-\theta^K_Y\,,
\qquad \vtd=A\lambda^3 R_t\,,\\
\bar P_c\left(Y_K\right) \equiv \left(1-\frac{\lambda^2}{2}\right)P_c\left(Y_K\right)\,,
\end{gather}
with $P_c\left(Y_K\right)=0.113\pm 0.017$
\cite{Gorbahn:2006bm}. Here 
$-\beta$ and $-\beta_s$ are the phases of $V_{td}$ and $-V_{ts}$, respectively.

In contrast to the remaining decays discussed in this paper and in \cite{Blanke:2006eb}, the {SD}
contribution calculated here is only a part of a dispersive contribution
to $K_L\to\mu^+\mu^-$ that is by far dominated by the absorptive contribution
with two internal photon exchanges. Consequently the SD contribution
 constitutes only a small fraction of the 
branching ratio. Moreover, because of long distance contributions to the
dispersive part of $K_L\to\mu^+\mu^-$, the extraction of the {SD}
part from the data is subject to considerable uncertainties. The most recent
estimate gives \cite{Isidori:2003ts}
\be\label{eq:KLmm-bound}
Br(K_L\to\mu^+\mu^-)_{\rm SD} \le 2.5 \cdot 10^{-9}\,,
\ee
to be compared with $(0.8\pm0.1)\cdot 10^{-9}$ in the SM 
\cite{Gorbahn:2006bm}.

\newsection{Numerical Analysis}\label{sec:num}

\begin{table}[ht]
\renewcommand{\arraystretch}{1}\setlength{\arraycolsep}{1pt}
\center{\small\begin{tabular}{|l|l|}
\hline
$\lambda=|V_{us}|= 0.2258(14)$ & $G_F= 1.16637\cdot 10^{-5}\gev^{-2}$ \qquad {} \\
$|V_{ub}| = 3.8(4)\cdot 10^{-3}$ &  $M_W = 80.398 \gev$ \\
$|V_{cb}|= 4.1(1)\cdot 10^{-2}$& $\alpha(M_Z) = 1/127.9$ \\
$\gamma = 78(12)^\circ $\hfill\cite{Bona:2006sa} & $\sin^2\theta_W = 0.23122$\\\cline{1-1}
$\Delta M_K= 0.5292(9)\cdot 10^{-2} \,\text{ps}^{-1}$ \qquad {} & $m_K^0= 497.614\mev$ \\
$|\eps_K|= 2.229(12)\cdot 10^{-3}$ \hfill\cite{Amsler:2008zz}& $m_{B_d}= 5279.5\mev$ \\\cline{1-1}
$\Delta M_d = 0.507(5) \,\text{ps}^{-1}$ & $m_{B_s} = 5366.4\mev$ \hfill\cite{Amsler:2008zz} \\\cline{2-2}
$\Delta M_s = 17.77(12) \,\text{ps}^{-1}$  & $\eta_1= 1.43(23)$ \hfill\cite{Herrlich:1993yv}\\\cline{2-2}
$S_{\psi K_S}= 0.672(24)$ \hfill\cite{Barberio:2008fa}&  $\eta_3=0.47(5)$ \hfill\cite{Herrlich:1995hh,Herrlich:1996vf} \\\hline
$\bar m_c = 1.27(2)\gev$ & $\eta_2=0.577(7)$ \\
$\bar m_t = 162.7(13)\gev$ \hfill\cite{Amsler:2008zz}& $\eta_B=0.55(1)$ \hfill \cite{Buras:1990fn,Urban:1997gw} \\\hline
$f_K = 156.1(8)\mev$ \hfill \cite{Antonelli:2008jg} & $f_{B_s} = 245(25)\mev$\\\cline{1-1}
$\hat B_K= 0.75(7)$ & $f_{B_d} = 200(20)\mev$ \\
$\hat B_{B_s} = 1.22(12)$ & $f_{B_s} \sqrt{\hat B_{B_s}} = 270(30)\mev$ \\
$\hat B_{B_d} = 1.22(12)$ & $f_{B_d} \sqrt{\hat B_{B_d}} = 225(25)\mev$ \\
$\hat B_{B_s}/\hat B_{B_d} = 1.00(3)$ \hfill \cite{Lubicz:2008am} & $\xi = 1.21(4)$ \hfill \cite{Lubicz:2008am}
 \\\hline
$m_e=0.5110\mev$ & $\tau(B_d)/\tau(B^+)=0.934(7)$\\
$m_\mu=105.66\mev$ & $\tau(B_s)=1.466(59)\,\text{ps}$ \\ 
$m_\tau=1.7770(3)\gev$ & $\tau(B^+)=1.638(11)\,\text{ps}$ \\
$\tau_\tau = 290.6(10)\cdot 10^{-3}\,\text{ps}$ \hfill\cite{Amsler:2008zz}& \hfill\cite{Amsler:2008zz}\\\hline
$F_8/F_\pi=1.28$  & $\theta_8=-22.2(18)^\circ$\\
$F_0/F_\pi=1.18(4)$\hfill\cite{Escribano:2005qq}& $\theta_0=-8.7(21)^\circ$ \hfill\cite{Escribano:2005qq}\\\hline
$Br(\mu \to e \gamma) < 1.2 \cdot 10^{-11}$ \hfill\cite{Brooks:1999pu}& $R(\mu\text{Ti} \to e\text{Ti}) < 4.3 \cdot 10^{-12}$\hfill\cite{Dohmen:1993mp}\\\hline
$Br(\mu^- \to e^- e^+ e^-) < 1.0 \cdot 10^{-12}$\quad\cite{Bellgardt:1987du} \\\cline{1-1}

\end{tabular}  }
\caption {\textit{Values of the experimental and theoretical
    quantities used as input parameters.} }
\label{tab:input}
\renewcommand{\arraystretch}{1.0}
\end{table}
The main purpose of the present flavour physics analysis is to identify the most evident LHT effects in both the quark and lepton sectors and, in particular, to study how our predictions of refs.~\cite{Blanke:2006eb,Blanke:2008ac,Blanke:2007db} are modified by the inclusion of the previously missed $\mathcal{O}(v^2/f^2)$ contribution to $Z^0$-penguin diagrams. 
This analysis also gives us the opportunity to update some of the input parameters, collected in Table  \ref{tab:input} together with the experimental constraints.
The input parameters are taken to be flatly distributed within their $1\sigma$ ranges, whereas the quark {flavour} observables     $\varepsilon_K$, $\Delta M_d$, $\Delta M_s$ and $S_{\psi K_S}$, resulting
from SM and LHT contributions, are required to lie within their experimental $1\sigma$
ranges, also shown in Table \ref{tab:input}. In the case of $\Delta M_K$ where the theoretical uncertainty is
large due to {poorly known} long-distance contributions, we allow the generated  
value to lie within $\pm 30\%$ of the experimental central value.
For the lepton sector, the experimental constraints used in the analysis are the upper bounds shown in Table  \ref{tab:input}.
All formulae for the observables used as constraints which were not discussed in the present paper can be found in our previous works \cite{Blanke:2006sb,Blanke:2006eb,Blanke:2008ac,Blanke:2007db}, with suitably corrected $Z^0$-penguin contributions {and adapted leptonic $U(1)_i$ charges} following Appendix \ref{sec:appA}.

As we are interested in pointing out potentially visible LHT effects, we do not consider specific scenarios for the LHT parameters, while we perform a general scan over the mirror fermion masses and the parameters of the mixing matrices $V_{Hd}$ and $V_{H\ell}$, with the NP scale $f$ fixed to 1 TeV and the mixing parameter of the T-even top Yukawa sector $x_L$ to 0.5, in agreement with EW precision tests.
In the quark sector a large number of points is generated where
mirror quark masses are varied in the interval [300\,GeV,1\,TeV], all angles of the $V_{Hd}$ matrix in the interval $[0,\pi/2]$, the phases between 0 and $2\pi$ and all SM input 
parameters are varied in their $1\sigma$ ranges. In the plots related to $K$ and $B$ physics we then show a sample of 9000 points that are consistent with all available $\Delta F=2$ constraints, as described above. {We do not specifically filter for
``interesting'' points, so that in Bayesian theory our method of randomly picking values for the model parameters corresponds to flat priors for these parameters in the respective
ranges in which the parameters were varied, see e.\,g.\ \cite{Trotta:2008bp}. Therefore the point density in the plots gives us an
idea of how likely it is for the LHT model to generate a certain effect. }
{The black point in the figures displays the SM predictions, while the light blue point originates from only the T-even contribution.}

For the study of LFV effects, we vary the mirror lepton masses in the interval [300\,GeV,1.5\,TeV] and the $V_{H\ell}$ angles and phases in the ranges $[0,\pi/2]$ and $[0,2\pi]$, respectively. In order to get a notion of what size of LFV effects can naturally be expected in the LHT model we show, following \cite{Blanke:2008zb}, density plots rather than scatter plots. The number of parameter points is large in the light orange areas, while it is small in the dark purple regions. 
In contrast to the quark sector the number of LFV decays for which useful experimental constraints exist is rather
limited. Basically only the constraints on $Br(\mu\to e\gamma)$, $Br(\mu^-\to
e^-e^+e^-)$ and $R(\mu\text{Ti}\to e\text{Ti})$  given in Table \ref{tab:input} can be
used. The situation may change significantly in the coming years and
the next decade mainly thanks to the measurements expected at the MEG {and J-PARC experiments}, at the LHC and possibly at a SuperB factory.
At present, with the available experimental upper bounds largely above the SM predictions, it is interesting to investigate how much the LHT effects in LFV observables can be enhanced by a smaller NP scale.
Therefore, we also perform a general scan with the NP scale set to $f= 500 \gev$, that is {approximately} the smallest value still allowed by EW precision tests.

\newsection{\boldmath Results for Rare $K$ and $B$ Decays}
\label{sec:KB}

As mentioned in the Introduction, all the decays that do not receive $Z^0$-penguin contributions are unaffected by
the $\mathcal{O}(v^2/f^2)$ contribution pointed out in \cite{Goto:2008fj,delAguila:2008zu}. Thus, in particular our LHT predictions for particle-antiparticle mixing and for the {branching ratio of the $B\to X_s\gamma$
decay} do not require to be revised.
Here we aim to present the results obtained for rare kaon and $B$ meson decays by including that $\mathcal{O}(v^2/f^2)$ contribution.

Before turning our attention to the updated predictions for rare $K$ and $B$ decays, we first address the question how naturally the existing $\Delta F=2$ data can be fulfilled in the LHT model. Since the data on $\eps_K$ puts the most stringent constraint on many extensions of the SM \cite{Bona:2007vi}, in particular on RS models \cite{Csaki:2008zd,Blanke:2008zb,Agashe:2008uz}, we focus on the fine-tuning required to fit these data. In order to allow for an easy comparison with the results obtained in the RS model with custodial protection \cite{Blanke:2008zb}, also in the present LHT analysis we use the Barbieri-Giudice measure $\Delta_\text{BG}(O)$ of fine-tuning \cite{Barbieri:1987fn} which quantifies the sensitivity of a given observable $O$ to small variations in the model parameters $p_j$ {($j=1,\dots,m$)}. It is defined by
\be
\Delta_\text{BG}(O)= \underset{j=1,\dots,m}{\text{max}} \{  \Delta_\text{BG}(O,p_j) \}\,,
\label{eq:ftBG}
\ee
with
\be
\Delta_\text{BG} (O,p_j) = \left| \frac{p_j}{O}\frac{\partial O}{\partial p_j}\right| \,,
\ee
where the normalisation factor $p_j/O$ appears in order not to be sensitive to the absolute size of $p_j$ and $O$.

\begin{figure}
\center{\includegraphics[width=.6\textwidth]{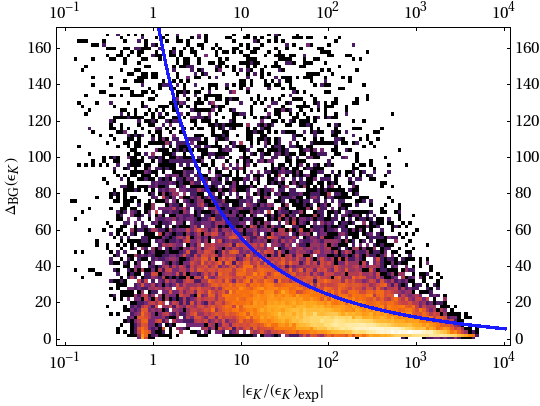}}
\caption{\it $\Delta_\text{BG}(\eps_K)$ as a function of $|\eps_K/(\eps_K)_\text{exp}|$. The solid blue line shows the {fine-tuning that is required on average} as a function of $\eps_K$.
\label{fig:epsKtun-noconst}}
\end{figure}
In Fig.\ \ref{fig:epsKtun-noconst} we show the measure $\Delta_\text{BG}(\eps_K)$ as a function of $\eps_K$, normalised to the data and plotted on a logarithmic axis. To this end we fixed the NP scale $f$ to {$1\tev$}, but did not impose any of the available flavour constraints. We observe that {generally} $\eps_K$ generated by the LHT dynamics is by roughly two orders of magnitude too large.
{Interestingly, a similar result is found in the custodially protected RS model \cite{Csaki:2008zd,Blanke:2008zb,Agashe:2008uz}.
This phenomenological similarity, however, hides different physics effects in the two models.
In the LHT model, where neither new tree-level effects nor new operators appear in addition to the SM ones, the large corrections to $\eps_K$ are a consequence of the arbitrary flavour structure. In the RS model, instead, new tree-level effects and new chirally enhanced operators are present and contribute to $K^0-\bar K^0$ mixing.}
The difference between these two models becomes clearer when studying the fine-tuning $\Delta_\text{BG}(\eps_K)$ needed to fit the data on $\eps_K$. While in the case of the custodially protected RS model the {fine-tuning required on average} increases quickly with decreasing $\eps_K$ \cite{Blanke:2008zb}, this increase is much slower in the LHT model, so that the {fine-tuning necessary} in order to fit the data on $\eps_K$ is much smaller in the LHT model than in the RS model. In particular in the LHT model we observe an increased density of points around $\eps_K\sim(\eps_K)_\text{exp}$ with no significant fine-tuning.
{The main reason for this difference is that in the LHT model, in spite of the arbitrary flavour structure, the new contributions are loop-suppressed, at variance with the RS model, and that in the LHT model the chosen value for the NP scale ($f=1 \tev$) is roughly twice the minimal value required by EW precision tests, while in the RS model the choice $f\simeq 1 \tev$ is {close to the lower bound}.}

\begin{figure}
\begin{minipage}{7.6cm}
\center{\includegraphics[width=\textwidth]{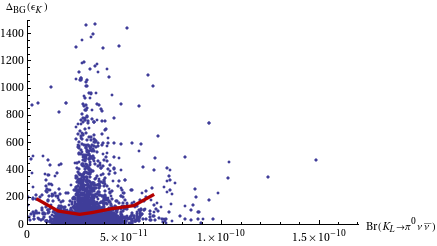}}
\end{minipage}\hfill
\begin{minipage}{7.6cm}
\center{\includegraphics[width=\textwidth]{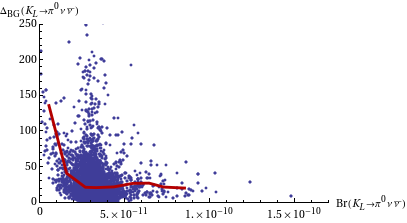}}
\end{minipage}
\caption{\it Left: $\Delta_\text{BG}(\eps_K)$ as a function of $Br(\klpn)$, showing only those points that fulfil all $\Delta F=2$ constraints. {The red line shows the on average required $\Delta_\text{BG}(\eps_K)$. Right: $\Delta_\text{BG}(\klpn)$ as a function of $Br(\klpn)$, showing only those points that fulfil all $\Delta F=2$ constraints. The red line shows the on average required $\Delta_\text{BG}(\klpn)$.
\label{fig:FineTuning-KL}}}
\end{figure}
In order to quantify how much fine-tuning is needed to fulfil the experimental constraint on $\eps_K$, we show in {the left panel of} Fig.\ \ref{fig:FineTuning-KL} $\Delta_\text{BG}(\eps_K)$ as a function of $Br(\klpn)$, showing only those points that fulfil all $\Delta F=2$ constraints, in particular the $\eps_K$ one. We observe that while certain regions of the LHT parameter space appear to be extremely fine-tuned, it is also possible to fulfil the existing constraints without large fine-tuning. In addition we notice that the required fine-tuning in $\eps_K$ is essentially uncorrelated to large effects in rare kaon decays, such as $\klpn$, so that large deviations from the SM in the latter are not necessarily fine-tuned.
{It is also interesting to check whether large effects in $\klpn$ require a large amount of fine-tuning in this observable. As can be seen from the right panel of Fig.\ \ref{fig:FineTuning-KL}, where we show $\Delta_\text{BG}(\klpn)$ as a function of $Br(\klpn)$, this is not the case -- only a significant suppression of $Br(\klpn)$ below its SM expectation requires some fine-tuning. This result is in fact easy to understand: The BG measure is by definition sensitive to large suppressions of an observable below its natural value. Large effects in $Br(\klpn)$ are however naturally generated by the LHT dynamics, the challenge is to bring these effects in agreement with the existing constraints, in particular from $\eps_K$.}

Rare decays in the kaon system represent privileged modes to search for NP, due to the strong CKM suppression present in the SM, which in NP models beyond MFV, like the LHT model, can be overcome by new sources of flavour violation.
\begin{figure}
\center{\includegraphics[width=.6\textwidth]{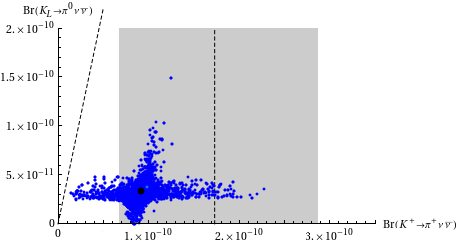}}
\caption{\it $Br(\klpn)$ as a function of $Br(\kpn)$. The shaded
    area represents the experimental central value and $1\sigma$-range for $Br(\kpn)$ and the
    GN-bound is displayed by the {dashed} line.}
\label{fig:KLKp}
\end{figure}
In Fig.\ \ref{fig:KLKp} we show the correlation between $Br(\kpn)$ and 
$Br(\klpn)$ as obtained from the general scan over the LHT parameters. The experimental
$1\sigma$-range for $Br(\kpn)$ \cite{Adler:2008zz} and the
model-independent Grossman-Nir (GN) bound \cite{Grossman:1997sk} are also
shown. We observe that the two
branches of possible points observed in \cite{Blanke:2006eb} are still present and that large enhancements with respect to the SM predictions are still allowed, though reduced by a factor {of $2-3$}. The first branch, which is parallel to the GN-bound, leads to possible huge enhancements in $Br(\klpn)$ so that values
as high as $1.5\cdot 10^{-10}$ are possible, being at the same time
consistent with the measured value for $Br(\kpn)$. 
On the second branch, which corresponds
to values for $Br(\klpn)$ rather close to its SM prediction,
 $Br(\kpn)$ is allowed to vary in the range $[1\cdot
10^{-11},2.5\cdot 10^{-10}]$. The presence of the two branches is a remnant of the specific operator structure of the LHT model and has been analysed in a model-independent manner in \cite{Blanke:2009pq}. Consequently observing one day the $K\to\pi\nu\bar\nu$ branching ratios outside these two branches would not only rule out the LHT model but at the same time put all models with a similar flavour structure in difficulties. On the other hand in models like the custodially protected RS model in which new flavour violating operators are present, no visible correlation is observed, so that an observation of the $K\to\pi\nu\bar\nu$ modes outside the two branches can be explained in such kind of models \cite{Blanke:2008yr}.

\begin{figure}
\center{\includegraphics[width=.6\textwidth]{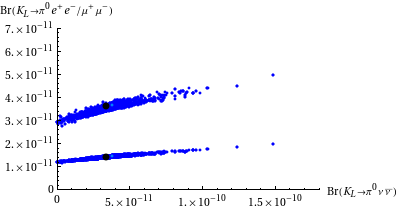}}
\caption{\it  $Br(K_L \to \pi^0
  \mu^+\mu^-)$ (lower curve) and $Br(K_L \to \pi^0
  e^+e^-)$ (upper curve) as functions of $Br(\klpn)$. }
\label{fig:KemuKp}
\end{figure}
In Fig.\ \ref{fig:KemuKp}, the correlation between the branching ratios of the CP-violating decays $K_L\to\pi^0 \ell^+\ell^-$ and $\klpn$ is shown. 
As in ref. \cite{Blanke:2006eb}, we find a strong correlation between the $K_L\to \pi^0\ell^+\ell^-$ and 
$\klpn$ decays, which is expected to be valid also beyond the LHT model, at least
in models in which no scalar operators contribute to $K_L\to\pi^0 \ell^+\ell^-$ {\cite{Isidori:2004rb,Friot:2004yr,Mescia:2006jd}}. In particular the observed correlation has also been found in the RS model with custodial protection \cite{Blanke:2008yr}.
The enhancement of $Br(K_L\to\pi^0 \ell^+\ell^-)$ in the LHT model with respect to the SM can be large, though smaller by approximately a factor of two than our finding in \cite{Blanke:2006eb}.
Values as high as $Br(K_L\to\pi^0 \mu^+\mu^-)=2\cdot 10^{-11}$ and $Br(K_L\to\pi^0 e^+e^-)=5\cdot 10^{-11}$ are allowed, but values larger than $1.7\cdot 10^{-11}$ and $4.5\cdot 10^{-11}$, respectively, require some tuning of the LHT parameters.  
 We note that a large enhancement of $Br(\klpn)$ automatically implies significant 
enhancements of $Br(K_L\to \pi^0\ell^+\ell^-)$ and that
different models, at least with the same operators as the SM, can then be distinguished
by their position along the correlation curve.
Moreover, measuring $Br(K_L\to\pi^0 \ell^+\ell^-)$ should 
allow for a rather precise prediction of $Br(\klpn)$.\enlargethispage{2mm}

\begin{figure}
\center{\includegraphics[width=.6\textwidth]{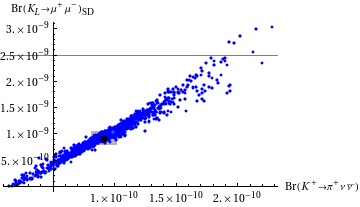}}
\caption{\it  $Br(K_L \to 
  \mu^+\mu^-)_\text{SM}$ as a function of $Br(\kpn)$. }
\label{fig:KLmuKp}
\end{figure}
We point out here a further interesting correlation that is found in the LHT model between the branching ratios of the CP-conserving decays  $K_L \to \mu^+ \mu^-$ and $\kpn$, as shown in Fig.\ \ref{fig:KLmuKp}. As discussed recently in \cite{Blanke:2008yr} this linear correlation should be contrasted with the inverse correlation between the two decays in question found in the custodially protected RS model. The origin of this difference is the operator structure of the models in question: While in the LHT model rare $K$ decays are mediated as in the SM by left-handed currents, in the RS model in question the flavour violating $Z$ coupling to right-handed quarks dominates.
In the LHT model consequently a large enhancement of $Br(\kpn)$ automatically implies a significant 
enhancement of $Br(K_L \to \mu^+ \mu^-)_\text{SD}$, which can be as high as $3 \cdot 10^{-9}$.
Values larger than the experimental upper bound $Br(K_L \to \mu^+ \mu^-)_\text{SD}<2.5 \cdot 10^{-9}$, displayed by the solid line in Fig.\ \ref{fig:KLmuKp}, turn out to be possible only with some parameter tuning.

Though the CKM suppression in rare kaon decays makes the $K$ system a particularly advantageous environment to look for NP effects, also the $B_{d,s}$ systems certainly deserve great attention, mainly so the clean rare decays $B_{d,s}  \to \mu^+ \mu^-$ and the phase of $B_s^0 - \bar B_s^0$ mixing.
The latter observable does not receive contribution from the $Z^0$-penguin and therefore from the  previously missed $\mathcal{O}(v^2/f^2)$ term, nevertheless we wish to update its LHT prediction here since the data \cite{Aaltonen:2007he,:2008fj,Brooijmans:2008nt} hints towards a possibly sizable deviation  
 \cite{Lenz:2006hd,Bona:2008jn} from the tiny SM value.
On the other hand,  $Br(B_{d,s}  \to \mu^+ \mu^-)$ are affected by that $\mathcal{O}(v^2/f^2)$ contribution which is included here.

\begin{figure}
\center{\includegraphics[width=.6\textwidth]{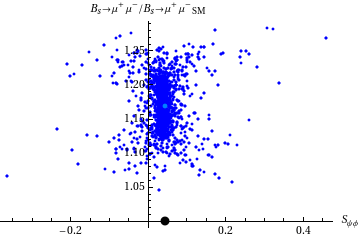}}
\caption{\it $Br(B_s\to\mu^+\mu^-)$, normalised to its SM value, as a
  function of $S_{\psi \phi}$.}
\label{fig:BsS}
\end{figure}
In Fig.\ \ref{fig:BsS} we show the ratio $Br(B_s\to\mu^+\mu^-)/Br(B_s\to\mu^+\mu^-)_\text{SM}$ as a function of $S_{\psi\phi}$. We observe that $Br(B_s\to\mu^+\mu^-)$ can be enhanced by at most 30\% over its SM value, and is bounded from below by the SM prediction. Observing a significant enhancement or a suppression of $Br(B_s\to\mu^+\mu^-)$ with respect to the SM would thus put the LHT model in severe difficulties. The CP-asymmetry $S_{\psi\phi}$ can vary in the range $[-0.4,+0.5]$, though values larger than 0.2 are quite unlikely. This means that the experimental value $0.2\le S_{\psi\phi} \le 1.0 \quad (95\%\text{ C.L.})$, obtained by the UTfit collaboration  \cite{Bona:2008jn,Bona:2006sa}
by combining the CDF \cite{Aaltonen:2007he} and D{\O} \cite{:2008fj} data\footnote{We note that a $2.2\sigma$ discrepancy with the SM prediction is obtained by the HFAG collaboration \cite{Barberio:2008fa}.}, 
can be explained within the LHT model, though with some tuning of the LHT parameters.

Very interesting is the golden relation between $Br(B_{d,s}  \to \mu^+ \mu^-)$ and the mass differences $\Delta M_{d,s}$, which can be written as \cite{Buras:2003td}
 \be\label{eq:r}
\frac{Br(B_s\to\mu^+\mu^-)}{Br(B_d\to\mu^+\mu^-)}= \frac{\hat
B_{B_d}}{\hat B_{B_s}} \frac{\tau(B_s)}{\tau(B_d)} \frac{\Delta
M_s}{\Delta M_d}\,r\,.
\ee
{The parameter $r$ is} equal to unity in constrained MFV (CMFV) \cite{Buras:2000dm,Buras:2003jf,Blanke:2006ig} models, i.\,e.\ in the absence of new sources of flavour violation and with only the SM operators present, whereas it can be generally different from unity.
\begin{figure}
\center{\includegraphics[width=.6\textwidth]{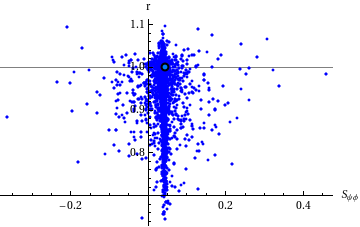}}
\caption{\it $r$ as a
  function of $S_{\psi \phi}$.}
\label{fig:rphis}
\end{figure}
In Fig.\ \ref{fig:rphis} we show the ratio $r$ as a
function of the CP asymmetry $S_{\psi\phi} $.
We observe that $r$ varies in the range $0.5\le r \le 1.1$, showing that visible deviations from the SM and CMFV prediction $r=1$ are allowed. We also find that $r<1$ is favoured over $r>1$ and that the largest departures from CMFV are found for SM-like values of $S_{\psi\phi}$.

\begin{figure}
\center{\includegraphics[width=.6\textwidth]{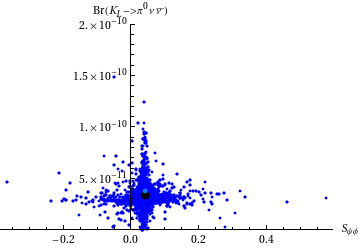}}
\caption{\it $Br(\klpn)$ as a
  function of $S_{\psi \phi}$.}
\label{fig:KLS}
\end{figure}
A further interesting question we want to address is whether large LHT effects can be simultaneously found in the $K$ and $B_{d,s}$ systems.
We consider the very clean and NP sensitive observables $Br(\klpn)$ and $S_{\psi\phi}$, whose correlation is shown in Fig.\ \ref{fig:KLS}.
From the clear pattern where two branches of points show up we learn that simultaneous deviations form the SM are unlikely but not impossible. A similar behaviour is found for the correlation between $Br(\kpn)$ and $S_{\psi\phi}$.

\begin{figure}
\center{\includegraphics[width=.6\textwidth]{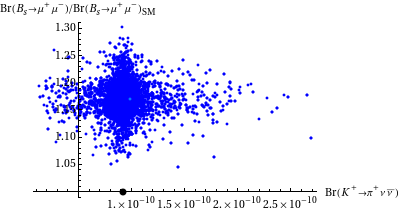}}
\caption{\it $Br(B_s\to\mu^+\mu^-)/Br(B_s\to\mu^+\mu^-)_\text{SM}$ as a
  function of $Br(\kpn)$.}
\label{fig:K+Bs}
\end{figure}
In Fig.\ \ref{fig:K+Bs} we show the correlation between $Br(B_s\to\mu^+\mu^-)$ and $Br(\kpn)$. A cross-like structure is also observed in this case, stemming again from the fact that the T-odd sector is unlikely to yield sizable contributions to both decays simultaneously. Due to the T-even contribution, displayed by the light blue point in the plot, a $\sim 15\%$ enhancement of $Br(B_s\to\mu^+\mu^-)$ with respect to the SM value can thus be expected in the LHT model if $Br(\kpn)$ deviates significantly from the SM.

\begin{figure}
\center{\includegraphics[width=.6\textwidth]{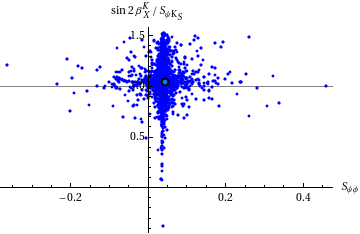}}
\caption{\it $\sin2\beta_X^K/S_{\psi K_S}$ as a
  function of $S_{\psi \phi}$.}
\label{fig:betaS}
\end{figure}
Finally in Fig.~\ref{fig:betaS} we show the ratio of $\sin2\beta$ extracted from the $K\to\pi\nu\bar\nu$ decays and from the mixing induced CP-asymmetry in $B_d\to\psi K_S$. In the SM and in MFV models this ratio equals unity \cite{Buchalla:1994tr,Buras:2001af}, but as seen in Fig.\ \ref{fig:betaS} the presence of non-MFV interactions in the LHT model allows for values different from unity, in particular {if $S_{\psi\phi}$ is SM-like}.

\newsection{Results for LFV Decays}\label{sec:LFV}
Among the LFV decays that we studied in \cite{Blanke:2007db}, those that receive contributions from the $Z^0$-penguin and thus require to be reanalysed with the inclusion of
the $\mathcal{O}(v^2/f^2)$ term, are $\ell_i^- \to \ell_j^- \ell_j^+ \ell_j^-$, $\tau^- \to \ell_i^- \ell_k^+ \ell_k^-$, $\tau \to \ell \pi (\eta, \eta^\prime)$ and the $\mu - e$ conversion {rate} in nuclei.
We present here the expectations for these observables in the LHT model, investigating whether the {maximal values allowed by the LHT model} turn out to be modified with respect to  \cite{Blanke:2007db}.

\begin{figure}
\center{\includegraphics[width=0.6\textwidth]{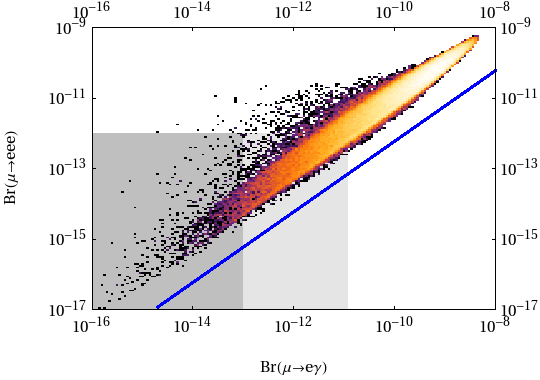}}
\caption{\it  Correlation between $\mu\to e\gamma$ and $\mu^-\to e^-e^+e^-$ as obtained from a general scan over the LHT parameters. The shaded area represents the present (light) and future (darker) experimental constraints. The solid blue line represents the dipole contribution to $Br(\mu^-\to e^-e^+e^-)$.}
\label{fig:megm3e}
\end{figure}

In Fig.~\ref{fig:megm3e} we show the correlation between $Br(\mu\to e\gamma)$ and $Br(\mu^-\to e^-e^+e^-)$ together with the experimental bounds on these decays.
We observe that the inclusion of the previously left out $\mathcal{O}(v^2/f^2)$ contribution does not spoil the two main results.
First, the great majority of points is outside the {experimentally} allowed range shown by the {light} grey area, which is even more obvious when considering the present density plot.
This implies, in agreement with the findings in \cite{Blanke:2007db,delAguila:2008zu}, that the $V_{H\ell}$ matrix must be much more hierarchical than $V_\text{PMNS}$ in order to satisfy the present upper bounds on $\mu\to e\gamma$ and $\mu^-\to e^-e^+e^-$.
{Secondly}, a strong correlation between $Br(\mu\to e\gamma)$ and $Br(\mu^-\to e^-e^+e^-)$ is observed, {stemming} from the common combination of $V_{H\ell}$ elements involved in these two decays.
We emphasise that the strong correlation between $Br(\mu\to e\gamma)$ and
$Br(\mu^-\to e^-e^+e^-)$ in the LHT model is not a common feature of all
extensions of the SM in which the structure of $\mu^-\to e^-e^+e^-$ is
generally much more complicated than in the LHT model. In particular {the LHT structure} differs significantly from models like the MSSM in which the dipole operator, displayed by the blue line, yields the dominant contribution to $Br(\mu^-\to e^-e^+e^-)$ \cite{Ellis:2002fe,Brignole:2004ah}. It is clear from
Fig.~\ref{fig:megm3e} that an improved upper bound on $\mu\to e\gamma$, which should be available from the MEG experiment
in the next years (shown by the dark grey area in Fig.~\ref{fig:megm3e}), and in particular its discovery will {provide important information}  on  $\mu^-\to e^-e^+e^-$ within the model in question. 

\begin{figure}
\center{\includegraphics[width=0.6\textwidth]{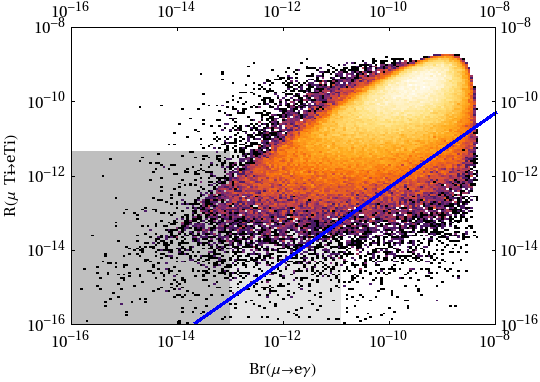}}
\caption{\it  Correlation between $\mu\to e\gamma$ and $\mu\to e$  conversion in Ti as obtained from a general scan over the LHT parameters. The shaded area represents the present (light) and future (darker) experimental constraints. The solid blue line represents the dipole contribution to $R(\mu \text{Ti}\to e\text{Ti})$.}
\label{fig:megeconv}
\end{figure}
Next in Fig.\ \ref{fig:megeconv} we show the $\mu\to e$ conversion rate in titanium (Ti), as a function of $Br(\mu\to e\gamma)$. We observe that the correlation between these two modes is much weaker than the one between $\mu\to e\gamma$ and $\mu^-\to e^-e^+e^-$. Consequently, the ratio of these two rates may again differ significantly from the prediction obtained in models where the dipole operator is dominant. {Such a distinction is however} not possible for some regions of the LHT parameter space, where the a priori dominant $Z^0$-penguin and box contributions cancel due to a destructive interference in $R(\mu \text{Ti}\to e\text{Ti})$.

\begin{figure}
\center{\includegraphics[width=0.6\textwidth]{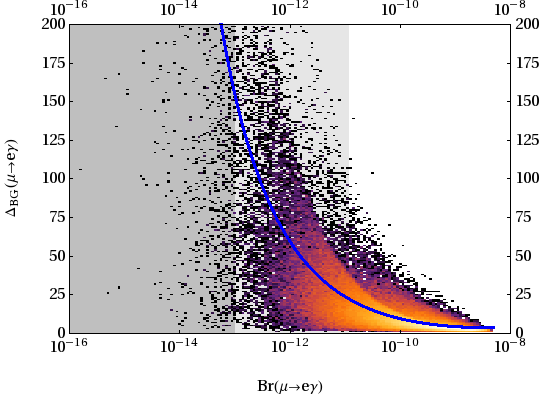}}
\caption{\it  Fine-tuning $\Delta_\text{BG}(\mu\to e\gamma)$ as a function of $Br(\mu\to e\gamma)$. {The blue curve shows the on average required fine-tuning as a function of $Br(\mu\to e\gamma)$.} The shaded area represents the present (light) and future (darker) experimental constraints.}
\label{fig:megtun}
\end{figure}
In order to quantify how naturally a suppression of the $\mu\to e\gamma$ decay rate below the present experimental bounds can be obtained, we consider how much fine-tuning is necessary to fulfil this bound. {We would like to remind the reader that the measure of fine-tuning $\Delta_\text{BG}$ defined in (\ref{eq:ftBG}) indicates the sensitivity of a particular observable with respect to a small change in the model parameters. It by no means allows to make statements for instance about the structure of the mixing matrices or the mass spectrum of the model, but only about how rapidly an observable changes in the neighborhood of a particular parameter configuration. No more than that the BG fine-tuning indicates whether or not one or more model parameters are accidentally small. From this it follows that a small BG fine-tuning does not preclude a hierarchical mixing matrix $V_{H\ell}$ or a degenerate mirror lepton mass spectrum.}
In Fig.\ \ref{fig:megtun} we show $\Delta_\text{BG}(\mu\to e\gamma)$ as a function of $Br(\mu\to e\gamma)$. We observe that for values of $Br(\mu\to e\gamma)\gsim 10^{-10}$ that are  already excluded by the data, the fine-tuning is generally small, $\Delta_\text{BG}(\mu\to e\gamma) < 25$. While with decreasing $Br(\mu\to e\gamma)$ large fine-tuning can be found, even for $Br(\mu\to e\gamma)\lsim 10^{-11}$ in agreement with the present experimental bound, a significant number of parameter points exist that are not subject to relevant fine-tuning. The situation will however change drastically if the MEG experiment pushes the bound on $Br(\mu\to e\gamma)$ further down to $\sim 10^{-13}$. We can see that only very specific LHT parameter points predict such small LFV effects in $\mu\to e$ transitions, so that the LHT model without any additional flavour symmetries would then be in difficulties to accommodate the data.

\begin{figure}
\center{\includegraphics[width=0.6\textwidth]{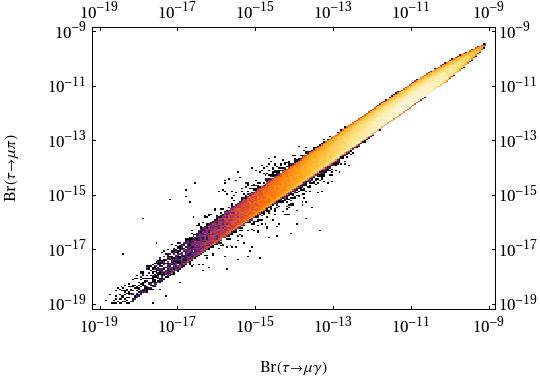}}
\caption{\it  $Br(\tau\to \mu\pi)$  as a function of $Br(\tau\to\mu\gamma)$.\label{fig:tmgtmp}}
\end{figure}

In Fig.~\ref{fig:tmgtmp} we show $Br(\tau\to\mu\pi)$  as a function of $Br(\tau\to\mu\gamma)$, imposing the constraints from $\mu\to e\gamma$ and $\mu^-\to e^-e^+e^-$. We find that $Br(\tau\to\mu\pi)$ can be as large as $4\cdot 10^{-10}$ which is smaller than the maximal value in  \cite{Blanke:2007db} by approximately a factor of five.
Similarly, the inclusion of the $\mathcal{O}(v^2/f^2)$ term reduces the LHT maximal values for $Br(\tau\to\mu\eta)$ and $Br(\tau\to\mu\eta')$ by about a factor of five which now turn out to be $\lsim 2\cdot 10^{-10}$ and $\simle 1\cdot 10^{-10}$, respectively.
Completely analogous {maximal values} and correlations can be found also for the decays $\tau\to e\pi,e\eta,e\eta'$ and $\tau\to e\gamma$. 

In Table \ref{tab:bounds} we collect the LHT {maximal values} obtained for the analysed $\tau$ decay branching ratios, together with the corresponding expected sensitivities of a SuperB factory \cite{Bona:2007qt}.
For completeness we also include the branching ratios that are unaffected by the $Z^0$-penguin contribution. As in \cite{Blanke:2007db}, we give these bounds both for $f=1000\gev$ and $f=500\gev$, in order to see the strong dependence on
the scale $f$.

\begin{table}[t]
{
\begin{center}
\begin{tabular}{|c|c|c|c|}
\hline
decay & $f=1000\gev$ & $f=500\gev$ & SuperB sensitivity \\\hline\hline
$\tau\to e\gamma$ & $8\cdot 10^{-10}$  & ${2\cdot 10^{-8}}$ & ${2\cdot10^{-9}}$  \\
$\tau\to \mu\gamma$ & $8\cdot 10^{-10}$  &$2\cdot 10^{-8}$   &${2\cdot10^{-9}}$ \\
$\tau^-\to e^-e^+e^-$ & $1\cdot10^{-10}$  & ${2\cdot10^{-8}}$   & $2\cdot10^{-10}$ \\
$\tau^-\to \mu^-\mu^+\mu^-$ & $1\cdot10^{-10}$  & ${2\cdot10^{-8}}$   & $2\cdot10^{-10}$ \\
$\tau^-\to e^-\mu^+\mu^-$ & $1\cdot10^{-10}$ & ${2\cdot10^{-8}}$   & \\
$\tau^-\to \mu^-e^+e^-$ & $1\cdot10^{-10}$ & ${2\cdot10^{-8}}$  & \\
$\tau^-\to \mu^-e^+\mu^-$ & $6\cdot10^{-14}$ & ${1\cdot10^{-13}}$ & \\
$\tau^-\to e^-\mu^+e^-$ & $6\cdot10^{-14}$ &${1\cdot10^{-13}}$   &  \\
$\tau\to\mu\pi$ & $4\cdot10^{-10} $  & ${5\cdot10^{-8}}$  & \\
$\tau\to e\pi$ & $4\cdot10^{-10} $ & ${5\cdot10^{-8}}$   & \\
$\tau\to\mu\eta$ & $2\cdot10^{-10}$  & ${2\cdot10^{-8}}$  & $4\cdot10^{-10}$ \\
$\tau\to e\eta$ & $2\cdot10^{-10}$  & ${2\cdot10^{-8}}$  & $6\cdot10^{-10}$ \\
$\tau\to \mu\eta'$ & $1\cdot10^{-10}$ & ${2\cdot10^{-8}}$  & \\
$\tau\to e\eta'$ & $1\cdot10^{-10}$ & ${2\cdot10^{-8}}$   & \\\hline
\end{tabular}
\end{center}
}
\caption{\it  {Maximal values} on LFV $\tau$ decay branching ratios in the LHT model, for two different values of the scale $f$, after imposing the constraints on $\mu\to e\gamma$ and $\mu^-\to e^-e^+e^-$. Including also the constraint $R(\mu\text{Ti}\to e\text{Ti})<5\cdot10^{-12}$  does not affect the obtained bounds. The expected SuperB sensitivities are also given where available \cite{Bona:2007qt}.\label{tab:bounds}}
\end{table}
We first observe that the branching ratios for LFV $\tau$ decays with three leptons in the final state are lowered by almost an order of magnitude once the UV-cutoff dependence is cancelled. 
In spite of this quantitative difference, the property that the {maximal values} on $\tau$ decays, except for
$\tau^-\to\mu^-e^+\mu^-$ and $\tau^-\to e^-\mu^+e^-$, increase by almost two
orders of magnitude when lowering the scale $f$ down to $500\gev$, is not spoilt by the inclusion of the missing $\mathcal{O}(v^2/f^2)$ term.
Moreover, as already found in  \cite{Blanke:2007db}, the bounds on $\tau^-\to\mu^-e^+\mu^-$ and $\tau^-\to e^-\mu^+e^-$ are quite independent of the value of $f$, due to the fact that the present lepton constraints are only effective for $\mu\to e$ transitions. 
Comparing the bounds obtained in the LHT model with the expected SuperB sensitivities, we observe that LFV effects induced by the LHT model can be observed at a SuperB facility provided the NP scale is small, $f<1\tev$.

We now discuss the correlations between various branching ratios that have been pointed out in Section 13 of \cite{Blanke:2007db} as powerful observables in providing a clear signature of the LHT model.
A general advantage of studying correlations is that they are less parameter-dependent than individual branching ratios.
{For the correlations in question it has also been found \cite{Blanke:2007db} that in the LHT model they present a pattern that differs significantly from the one of the MSSM, in particular at large $\tan\beta$.
The main reason is that in {the LHT model} the dipole contributions to the decays $\ell_i^-\to \ell_j^-\ell_j^+\ell_j^-$ and $\ell_i^-\to \ell_j^-\ell_k^+\ell_k^-$ can be fully neglected in comparison to $Z^0$-penguin and box diagram contributions. In fact, the neutral gauge boson $(Z_H,A_H)$ contributions interfere destructively with the $W_H^\pm$ contributions to the dipole operator functions, but constructively in $Z^0$-penguin and box contributions.
As seen in Fig.\ \ref{fig:megm3e} where the dipole contribution is represented by the solid blue line, 
this pattern is still valid once the $\mathcal{O}(v^2/f^2)$ term is included in the $Z^0$-penguin.
However, the most important difference between these two models is the large $\tan\beta$ enhancement of dipole operators characteristic for the MSSM \cite{Hisano:2009ae}, which is absent in the LHT model. For
 $\tan\beta=1$ the distinction between these two models on the basis of LFV 
 would be much more difficult.\footnote{We thank Paride Paradisi for a useful discussion on this point.}}

\begin{table}
{\renewcommand{\arraystretch}{1.5}
\begin{center}
\begin{tabular}{|c|c|c|c|}
\hline
ratio & LHT  & MSSM (dipole) & MSSM (Higgs) \\\hline\hline
$\frac{Br(\mu^-\to e^-e^+e^-)}{Br(\mu\to e\gamma)}$  & \hspace{.8cm} 0.02\dots1\hspace{.8cm}  & $\sim6\cdot10^{-3}$ &$\sim6\cdot10^{-3}$  \\
$\frac{Br(\tau^-\to e^-e^+e^-)}{Br(\tau\to e\gamma)}$   & 0.04\dots0.4     &$\sim1\cdot10^{-2}$ & ${\sim1\cdot10^{-2}}$\\
$\frac{Br(\tau^-\to \mu^-\mu^+\mu^-)}{Br(\tau\to \mu\gamma)}$  &0.04\dots0.4     &$\sim2\cdot10^{-3}$ & $0.06\dots0.1$ \\\hline
$\frac{Br(\tau^-\to e^-\mu^+\mu^-)}{Br(\tau\to e\gamma)}$  & 0.04\dots0.3     &$\sim2\cdot10^{-3}$ & $0.02\dots0.04$ \\
$\frac{Br(\tau^-\to \mu^-e^+e^-)}{Br(\tau\to \mu\gamma)}$  & 0.04\dots0.3    &$\sim1\cdot10^{-2}$ & ${\sim1\cdot10^{-2}}$\\
$\frac{Br(\tau^-\to e^-e^+e^-)}{Br(\tau^-\to e^-\mu^+\mu^-)}$     & 0.8\dots2.0   &$\sim5$ & 0.3\dots0.5\\
$\frac{Br(\tau^-\to \mu^-\mu^+\mu^-)}{Br(\tau^-\to \mu^-e^+e^-)}$   & 0.7\dots1.6    &$\sim0.2$ & 5\dots10 \\\hline
$\frac{R(\mu\text{Ti}\to e\text{Ti})}{Br(\mu\to e\gamma)}$  & $10^{-3}\dots 10^2$     & $\sim 5\cdot 10^{-3}$ & $0.08\dots0.15$ \\\hline
\end{tabular}
\end{center}\renewcommand{\arraystretch}{1.0}
}
\caption{\it  Comparison of various ratios of branching ratios in the LHT model ($f=1\tev$) and in the MSSM without \cite{Ellis:2002fe,Brignole:2004ah} and with \cite{Paradisi:2005tk,Paradisi:2006jp} significant Higgs contributions.\label{tab:ratios}}
\end{table}

The ranges found for these ratios in the LHT model (for $f=1\tev$) once the UV-cutoff dependence is cancelled, are compared with the corresponding values in the MSSM in Table \ref{tab:ratios}, both in the case of dipole dominance and significant Higgs contributions.
We observe that the allowed ranges for the correlations in the LHT {model} shown in the first five lines of Table \ref{tab:ratios} turn out to be of the same order of magnitude but smaller than in  \cite{Blanke:2007db}.
It is important to stress that the measurement of these quantities could still allow for a clear distinction of the LHT model from the MSSM. We note that the obtained ranges depend only weakly on the scale $f$, so that a distinction along these lines is possible for any value of $f$.

\newsection{Conclusions}

We have presented an update of our 2006-2007 results for FCNC processes in the LHT {model}, including the previously missed $\mathcal{O}(v^2/f^2)$ contribution to $Z^0$-penguin diagrams and updating some {theoretical and experimental} inputs.
We have identified the most evident LHT effects in both the quark and lepton sectors and pointed out the decay channels that could allow for a clear distinction from other NP models.
The main results of this analysis are summarised below.

{While the data on $\eps_K$ provide a stringent constraint on the LHT parameter space, much less fine-tuning is needed to fulfil this constraint than in the RS model with custodial protection.}

In the kaon system large enhancements of the branching ratios $Br(\klpn)$, $Br(\kpn)$ and $Br(K_L\to\pi^0 \ell^+\ell^-)$ with respect to the SM predictions are possible.
Though the removal of the divergence reduces these enhancements by approximately a factor of two, the strong correlations among them are not modified {and provide a useful tool to distinguish the LHT model from other NP scenarios}.
Another interesting LHT correlation, which we have studied here for the first time, is between $Br(\kpn)$ and $Br(K_L \to \mu^+ \mu^-)$, pointing out that it is opposite and therefore distinguishable from the correlation predicted in the custodially protected RS model. Moreover, $Br(K_L \to \mu^+ \mu^-)_\text{SD}$ in {the LHT model} can be as large as $2.5 \cdot 10^{-9}$, that is much larger than the SM prediction.

In the $B$ system, the most interesting observable at present is the phase of $B_s^0-\bar B_s^0$ mixing. We have therefore updated the LHT prediction for the CP-asymmetry $S_{\psi \phi}$, though it is not affected by the $Z^0$-penguin contribution omitted in our previous analysis. We find that $S_{\psi \phi}$ can vary in the range $[-0.4, +0.5 ]$, so that the measured deviation from the SM can be explained in the LHT model, though with some tuning of the parameters, while larger values can easily be obtained in the RS model.

In the lepton sector we find that the inclusion of the previously left-out $\mathcal{O}(v^2/f^2)$ contribution does not spoil two important results, namely the requirement of a highly hierarchical $V_{H\ell}$ matrix in order to satisfy the present upper bounds on $\mu \to e \gamma$ and $\mu^- \to e^- e^+ e^-$ and the strong correlation between the branching ratios of these two decays.
Here we have also studied the fine-tuning required in the model in order to satisfy the present experimental bound on $Br(\mu \to e \gamma)$, finding that it is not relevant but that it would become important if the MEG experiment pushes the bound further down to $\sim10^{-13}$. {As mentioned in Section~\ref{sec:LFV} here we state again that a small fine-tuning as defined in (\ref{eq:ftBG}) does not exclude a hierarchical mixing matrix $V_{H\ell}$.}
The inclusion of the $\mathcal{O}(v^2/f^2)$ term turns out to {reduce  the LHT {maximal values} for the branching ratios of the decays $\tau \to \mu(e) \pi$, $\tau \to \mu(e) \eta$, $\tau \to \mu(e) \eta'$ by approximately a factor of five}, whose enhancements with respect to the SM are nevertheless large and could be visible at a SuperB factory.
Similarly, the branching ratios for LFV $\tau$ decays with three leptons in the final state are lowered by almost an order of magnitude once the UV-cutoff dependence is cancelled, but are still large enough to be observed at a SuperB facility provided the NP scale is small enough, $f < 1 \tev$.
An important feature that is not affected by the removal of the UV-cutoff dependence is the dominance of $Z^0$-penguin and box diagram contributions relative to the dipole contributions in the decays $\ell_i^- \to \ell_j^- \ell_j^+ \ell_j^-$ and $\ell_i^- \to \ell_j^- \ell_k^+ \ell_k^-$.
This LHT feature, being in contrast to the MSSM dipole dominance, allows for a distinction between these two models, in particular when looking at the ratios collected in Table~\ref{tab:ratios}.

\subsection*{Acknowledgements}
This research was partially supported by {the Cluster of Excellence `Origin and Structure of the Universe', the Graduiertenkolleg GRK 1054 of DFG and by} the German `Bundesministerium f{\"u}r Bildung und
Forschung' under contract 05HT6WOA.

\begin{appendix}

\newsection{Appendix}
\label{sec:appA}

\subsection{Formulae for Rare Decay Branching Ratios}

The modifications of some formulae in our papers~\cite{Blanke:2006eb},~\cite{Blanke:2007db} and~\cite{Blanke:2007wr}, which are required since there is no divergence in the $Z^0$-penguin contributions (see Section \ref{sec:cutoff}), can be summarised in a very compact manner. 
The divergent contribution $z_i\, S_\text{odd}$ should be replaced everywhere by a finite expression as follows
\be\label{eq:div}
z_i\, S_\text{odd} \to z_i \left( \frac{z_i^2 - 2\,z_i + 4}{(1 - z_i)^2}\log(z_i) + \frac{7 - z_i}{2\,(1-z_i)} \right)\,.
\ee
{Here $z_i = {m_H^i}^2/M_{W_H}^2$, with $m_H^i$ and $M_{W_H}$ being the mirror fermion and $W_H$ masses, respectively.}

The adaptation of $U(1)_i$ charges of leptons to the ones used by Goto et al. \cite{Goto:2008fj} as discussed in Section \ref{sec:cutoff} implies that the function $G_2$ in the functions $J^{\nu\bar\nu}$ and $J^{\mu\bar\mu}$ in (5.3) and (5.4) of \cite{Blanke:2006eb} and in $J^{u\bar u}$ and $J^{d\bar d}$ in (4.7) and (4.8) of \cite{Blanke:2007db} should be replaced by $-G_2$.

It should be emphasised that while the removal of the divergence by the authors of~\cite{Goto:2008fj,delAguila:2008zu} had a visible impact on our numerical results, the sign flip {of the function $G_2$} is numerically irrelevant (typically an $\ord (1\%)$ effect).

\subsection{Feynman Rules}

Below we list those LHT Feynman rules  that have to be modified with respect to \cite{Blanke:2006eb}.

The inclusion of the $v^2/f^2$ correction to the $Z_L$ and $W_L$ couplings of mirror fermions pointed out in \cite{Goto:2008fj,delAguila:2008zu} leads to the following modifications:
\setlength{\arraycolsep}{2pt}
\begin{center}
\begin{longtable}{|c|c|}  \hline
$\bar u^i_H Z_{L}^\mu u^i_H$ &
$\frac{ig}{\cos{\theta_W}}\left[\left(\frac{1}{2}-\frac{2}{3}\sin^2{\theta_W}\right)-\frac{v^2}{8f^2}P_R\right]\gamma^\mu$ 
\\\hline
$\bar u^i_{H} W^{+\mu}_{L} d^j_{H}$ & $\frac{ig}{\sqrt{2}}\left(1-\frac{v^2}{8f^2}P_R\right)
\delta_{ij} \gamma^\mu$ 
\\\hline
$\bar \nu^i_H Z_{L}^\mu \nu^i_H$ &
$\frac{ig}{\cos{\theta_W}} \left(\frac{1}{2}-\frac{v^2}{8f^2}P_R\right)
\gamma^\mu$ \\\hline 
$\bar \nu^i_{H} W^{+\mu}_{L} \ell^j_{H}$ &
$\frac{ig}{\sqrt{2}}\left(1-\frac{v^2}{8f^2}P_R\right)\delta_{ij} \gamma^\mu$ \\\hline
\end{longtable}
\end{center}

The adaptation of the $U(1)_i$ charges of leptons to the ones used in \cite{Goto:2008fj} leads to the following Feynman rules for $Z_H$ and $A_H$ couplings to leptons:
\begin{center}
\begin{longtable}{|c|c|}\hline
$\bar \nu^i_{H} A_{H}^\mu \nu^j$ & $\left(-\frac{ig'}{10}-\frac{ig}{2}x_H\frac{v^2}{f^2}
\right)(V_{H\nu})_{ij}\gamma^\mu P_L$\\\hline
$\bar \nu^i_{H} Z_{H}^\mu \nu^j$ & $\left(\frac{ig}{2}-\frac{ig'}{10}x_H\frac{v^2}{f^2}
\right)(V_{H\nu})_{ij}\gamma^\mu P_L$\\\hline
$\bar \ell^i_{H} A_{H}^\mu \ell^j$ & $\left(-
  \frac{ig'}{10}+\frac{ig}{2}x_H\frac{v^2}{f^2} \right)(V_{H\ell})_{ij}\gamma^\mu P_L$\\\hline
$\bar \ell^i_{H} Z_{H}^\mu \ell^j$ & $\left(-\frac{ig}{2}-\frac{ig'}{10}x_H\frac{v^2}{f^2} \right)(V_{H\ell})_{ij}\gamma^\mu
P_L$\\\hline
\end{longtable}
\end{center}

Finally the following typos occurred in the rules of \cite{Blanke:2006eb}:
\bi
\item
The PMNS matrix has to be replaced by its hermitian conjugate at all occurrences.
\item
The overall sign of the triple gauge boson vertices has to be reversed.
\ei

\end{appendix}

\providecommand{\href}[2]{#2}\begingroup\raggedright\endgroup


\end{document}